\documentclass[sigconf,natbib=true]{acmart}

\AtBeginDocument{%
  }

\setcopyright{acmlicensed}
\copyrightyear{2026}
\acmYear{2026}
\acmDOI{XXXXXXX.XXXXXXX}
\acmConference[Acronym 'XX]{Make sure to enter the correct
conference title from your rights confirmation email}{June 03--05,
2018}{Woodstock, NY}
\acmISBN{978-1-4503-XXXX-X/2018/06}

\usepackage{amsmath} 
\usepackage{amsfonts} 
\usepackage{mathrsfs}
\usepackage{enumitem}
\usepackage{booktabs}  
\usepackage{tcolorbox}
\usepackage{multirow} 
\usepackage{graphicx}  
\usepackage{amsmath}  
\usepackage{mathrsfs}  
\usepackage{subcaption}
\usepackage{caption}
\usepackage{subcaption}

\usepackage{algorithm}
\usepackage{algorithmic}

\begin{document}

\title{Learning to Reflect and Correct: Towards Better Decoding Trajectories for Large-Scale Generative Recommendation}


\author{Haibo Xing}
\orcid{0009-0006-5786-7627}
\authornotemark[1]
\affiliation{%
  \institution{Alibaba International Digital Commerce Group}
  \city{Hangzhou} 
  \state{} 
  \country{China}
}
\email{xinghaibo.xhb@alibaba-inc.com}

\author{Hao Deng}
\orcid{0009-0002-6335-7405}
\authornote{Contributed equally to this research.} 
\affiliation{%
  \institution{Alibaba International Digital Commerce Group}
   \city{Beijing} 
   \state{} 
   \country{China}
}
\email{denghao.deng@alibaba-inc.com}

\author{Lingyu Mu}
\affiliation{
  \institution{Alibaba International Digital Commerce Group}
  \city{Beijing} 
  \state{} 
  \country{China}
}
\email{moulingyu.mly@alibaba-inc.com}

\author{Jinxin Hu}
\orcid{0000-0002-7252-5207}
\authornote{Corresponding authors.}
\affiliation{
  \institution{Alibaba International Digital Commerce Group}
  \city{Beijing} 
  \state{} 
  \country{China}
}
\email{jinxin.hjx@alibaba-inc.com}

\author{Yu Zhang}
\orcid{0000-0002-6057-7886}
\affiliation{
  \institution{Alibaba International Digital Commerce Group}
  \city{Beijing} 
  \state{} 
  \country{China}
}
\email{daoji@lazada.com}

\author{Xiaoyi Zeng}
\orcid{0000-0002-3742-4910}
\affiliation{
  \institution{Alibaba International Digital Commerce Group}
  \city{Hangzhou} 
  \state{} 
  \country{China}
}
\email{yuanhan@taobao.com}

\author{Jing Zhang$^\dag$}
\orcid{0000-0001-6595-7661}
\affiliation{
  \institution{School of Computer Science, Wuhan University}
  \city{Wuhan} 
  \state{} 
  \country{China}
}
\email{jingzhang.cv@gmail.com}
\renewcommand{\shortauthors}{Trovato et al.}

\begin{abstract}
Generative Recommendation (GR) has become a promising paradigm for large-scale recommendation systems. However, existing GR models typically perform single-pass decoding without explicit refinement, causing early deviations to accumulate and ultimately degrade recommendation quality. To tackle this problem, we propose GRC, which is, to our knowledge, the first structured reflection-correction framework for GR that extends standard decoding into a \textbf{G}eneration–\textbf{R}eflection–\textbf{C}orrection (GRC) process. Concretely, GRC introduces a supervised reflection–correction template that decomposes the decoding process into initial draft generation, multi-granular reflection, and reflection-guided correction, thereby enabling structured reflection and correction in the semantic token space. To further explore the enlarged refinement space introduced by the GRC process, we optimize the entire GRC trajectory with GRPO-based reinforcement learning, under a carefully designed reward function with token-level and trajectory-level signals. For efficient online serving, we propose an Entropy-Guided Reflection Scheduling (EGRS) strategy that dynamically allocates more correction budget to high-uncertainty decoding trajectories during beam search. Extensive experiments on real-world datasets show that GRC consistently outperforms six state-of-the-art baselines by up to 15.74\%, and online A/B tests demonstrate its substantial practical value in large-scale industrial recommendation, delivering a 1.79\% lift in advertising revenue with only modest latency overhead.
\end{abstract}

\ccsdesc[500]{Information systems~Retrieval models and ranking}

\keywords{Generative Recommendation, Reflection, Reinforcement learning}


\maketitle

\section{Introduction}
\begin{figure}[htbp]
  \includegraphics[width=0.42\textwidth]{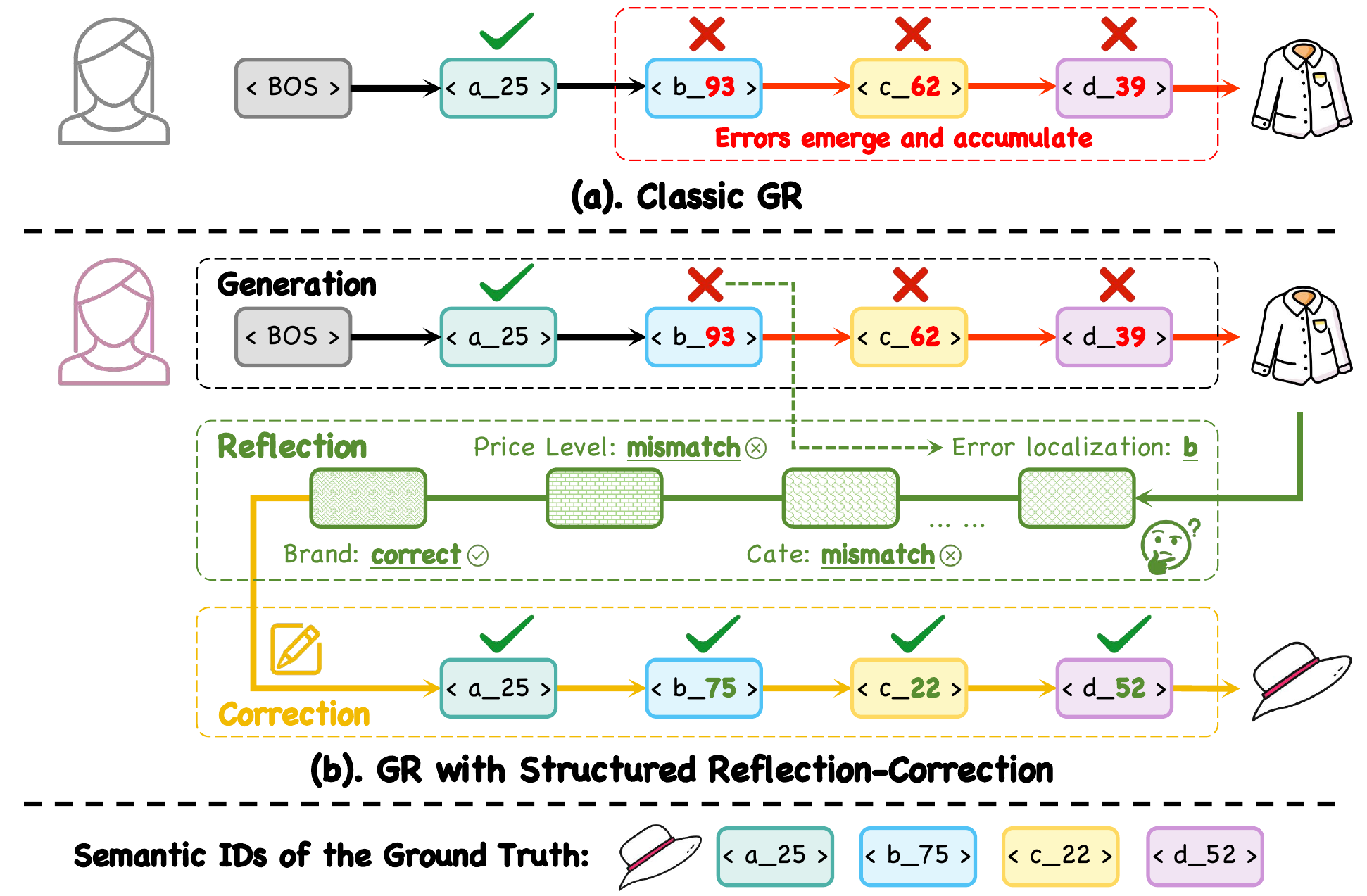}
  \vspace{-5pt}
  \caption{
  Illustration of Generative Recommendation. (a) Classic GR, where errors accumulate along one-pass autoregressive generation. (b) Our GRC, which augments generation with structured reflection and correction.
  }
  \Description{xx.}
  \label{fig:fig_intro}
  \vspace{-0.5cm} 
\end{figure}
In large-scale recommender systems, the retrieval stage plays a central role in matching user needs to massive item pools\ \citep{youtubeDNN,mind,ComiRec}. Recent advances in discrete semantic representations and sequence modeling have led to semantic-token-based Generative Recommendation (GR)\ \citep{yang2025sparse,rajput2023recommender,onerec}, which is increasingly adopted in the real-world retrieval stage. In this paradigm, item embeddings are first quantized into multi-level discrete tokens\ \citep{rqvae,ParallelID,mmq}, and the autoregressive GR then generates the target item’s semantic token trajectory directly in this discrete space, bypassing conventional dual‑tower retrieval\ \citep{lin2024enhancing,esans}. By sharing tokens across semantically similar items and enabling finer-grained preference modeling, token-driven GR shows strong potential for industrial applications such as e-commerce and content recommendation\ \citep{reg4rec, sid4rec}.

Despite these advantages, most GR methods still follow a \textbf{one-pass autoregressive} paradigm\ \citep{GRsurveyV1}: given user behavior, the model generates a complete semantic token sequence in a single pass and directly uses it for retrieval, without explicit inspection or refinement. In this setting, early decoding errors emerge and accumulate along the autoregressive chain\ \citep{GRsurveyV2}, resembling hallucination-like drift in large language models (LLMs) decoding\ \citep{HNeurons,FaithLens}. For multi-level semantic tokens, the first token typically selects a coarse semantic cluster, while later tokens specify finer-grained attributes\ \citep{rqvae}. Once an early token deviates, subsequent tokens can only specialize within the wrong cluster, causing the retrieval trajectory to drift from the target and limiting GR accuracy.

Fundamentally, existing methods rely on supervised imitation with teacher forcing to fit each item’s semantic token sequence. This confines training to ground-truth trajectories, but provides no explicit mechanism to evaluate or repair model-generated trajectories at inference. As a result, it is difficult to effectively mitigate error accumulation in the autoregressive generation process. Recent self-reflection mechanisms in LLMs \citep{selfRefine,Reflection} offer a promising way to mitigate error propagation by generating a critique and then refining the output accordingly. However, they typically rely on free-form natural-language reflections and explanations, which are hard to align with discrete semantic tokens in GR, hindering fine-grained error localization and correction. Moreover, multi-step natural-language generation introduces substantial latency overhead and unpredictable tail latency, making it impractical for real-time industrial deployment. To sidestep these issues, ReaRec \citep{rearec} performs implicit multi-step reasoning to refine item representations at inference time. However, this implicit refinement is inherently black-box and fails to support explicit supervision for correction, thereby undercutting the effectiveness of self-reflection mechanisms for targeted error diagnosis and correction in GR.

To address these challenges, we draw inspiration from self-reflection while deliberately avoiding free-form textual feedback. To this end, we propose \textbf{GRC}, a structured framework for GR that replaces conventional one-pass semantic-token generation with an explicit \textbf{G}eneration–\textbf{R}eflection–\textbf{C}orrection process. During supervised fine-tuning (SFT), we construct a \textbf{Structured Reflection–Correction Template} that specifies how these three components are serialized into a \emph{single} supervision trajectory. As illustrated in Figure~\ref{fig:fig_intro}(b), each training episode consists of three consecutive components: generation, reflection, and correction, forming a unified trajectory that endows the model with an initial self-correction capability. Concretely, in the generation component, the model first produces a candidate semantic-token trajectory as an initial retrieval draft, exactly as in standard GR. In the reflection component, instead of emitting free-form natural language, the model generates \emph{structured reflection signals} along two axes: (i) token-level error localization, identifying the first position in the draft where the generated token deviates from the reference trajectory and (ii) semantic-level consistency judgments (\textit{e.g.}, category and brand), indicating how the draft deviates from the reference. In the correction component, the model then rewrites the trajectory conditioned on both the draft and these structured reflection signals, and the corrected trajectory is used for retrieval. By following this template, GRC (i) explicitly instantiates generation, reflection, and correction at the semantic-token level, and (ii) converts self-reflection from opaque natural-language critiques into a structured signal space that can be directly supervised and optimized. This enables the model to learn to \emph{evaluate} and \emph{repair} off-target trajectories, rather than merely imitating ground-truth token sequences.

Incorporating GRC substantially enlarges the trajectory revision space, since varying reflection signals and correction edits can yield a \textbf{combinatorially} large set of generation–reflection–correction trajectories that simple heuristics cannot cover. If we rely only on the SFT stage, the model merely fits a small set of demonstration trajectories with fixed reflection–correction behaviors, leaving many potentially better strategies undiscovered. To address this, we further introduce a Reinforcement Learning (RL) stage based on Group Relative Policy Optimization (GRPO) \cite{shao2024deepseekmath} to enhance the model’s ability to explore self-correction policies in the enlarged decision space. We design episode structures and reward decompositions that are aligned with the GRC framework, so that the model learns a structured reflection–correction policy of "first identify where it went wrong, then fix it in a targeted way" rather than simply generating longer trajectories. In terms of reward design, we move beyond assessing only the correctness of the final generated sequence and introduce three fine-grained, reflection-oriented rewards that directly evaluate the quality of error localization and correction behavior. Through this RL process, the model learns stable self-correction strategies in the complex reflection–correction decision space, which in turn significantly improves the effectiveness and robustness of GR in large-scale recommendation scenarios.


Considering the real-time serving constraints in large-scale recommendation, the reflection–correction overhead incurred by GRC calls for selective application at inference time. We therefore propose Entropy-Guided Reflection Scheduling (EGRS), which leverages reflection-time uncertainty to prioritize candidates with higher correction potential, enabling more effective use of a fixed self-correction budget. Concretely, EGRS uses reflection-time token entropy as a lightweight proxy for correction potential and, under a fixed beam budget, biases beam selection toward more uncertain trajectories for further refinement. This design preserves the standard beam-search interface and keeps inference latency practical.

Our main contributions are as follows:
\begin{itemize}[noitemsep, topsep=0pt, leftmargin=*]
    \item We propose GRC, an explicit self-correction framework for GR. To the best of our knowledge, it is the first approach to move beyond one-pass decoding by modeling a structured Generation–Reflection–Correction process in the semantic-token space. By reducing error propagation in autoregressive semantic-token generation, GRC yields more robust trajectories and consistently improves retrieval accuracy.

    \item We further enhance GRC’s self-reflection and self-correction capabilities by introducing a GRPO-based reinforcement learning stage. In this stage, GRC directly optimizes reflection–correction policies in the expanded trajectory space, using an episode formulation and a reward design tailored to GRC. For efficient deployment, we further propose EGRS, which prioritizes trajectories with higher uncertainty and correction potential, thereby better realizing the benefits of reflection under strict latency constraints.

    \item We conduct extensive offline experiments on real-world datasets and deploy GRC in an online advertising system. The results show that our framework achieves consistent improvements over strong baselines and is practical for large-scale industrial deployment, with only a modest increase in inference latency.

\end{itemize}

\section{Related Work}
\subsection{Generative Recommendation}
In large-scale recommender systems, retrieval is typically implemented with dual-tower models that embed users and items into a shared space and retrieve candidates via Approximate Nearest Neighbor (ANN) search~\cite{zheng2022multi,esans,deng2025heterrec}. Despite their effectiveness, these models rely on item-ID–specific parameters, exacerbating popularity bias in million-scale catalogs and limiting generalization and semantic utilization~\cite{zhai2024actions}. To mitigate this issue, recent work explores GR with discrete semantic tokens. Early work such as TIGER \cite{rajput2023recommender} uses RQ-VAE to encode item content into multi-level semantic tokens, and then trains an autoregressive Transformer to generate the target item’s token sequence. Later work improves tokenization and codebook design for higher expressiveness~\cite{deng2025onerec,mmq,ParallelID}. Most GR models are trained with teacher forcing but decoded in one pass from self-generated prefixes, leading to exposure bias and error propagation \cite{pozzi2025mitigating,yang2025sparse,zhai2024actions,rajput2023recommender}. Early token errors shift later conditionals and cause the decoded sequence to drift in the item space, degrading retrieval quality. A few studies introduce limited correction mechanisms for GR \cite{yang2025sparse,yang2024unifying}. For example, COBRA~\cite{yang2025sparse} augments discrete tokens with continuous vectors and alternates between token generation and dense embedding prediction, leveraging ANN retrieval to partly mitigate early-token errors. ReaRec~\cite{rearec} introduces an inference-time refinement procedure that iteratively updates item representations for GR. However, these mechanisms remain largely implicit: they neither explicitly localize errors nor revise previously generated tokens along the trajectory.

In comparison, GRC introduces an explicit self-correction module for GR that localizes token-level errors and checks semantic consistency during decoding, revising the semantic token sequence accordingly and thereby reducing error propagation.

\subsection{Self-Reflection in Large Language Models}
Reasoning and reflection have been central to performance gains of LLMs on complex tasks \cite{manning2022human,wu2024large}. Many methods replace one-pass generation with multi-round reason-reflect-revise procedures. On the reasoning side, models construct structured reasoning paths, such as Tree of Thoughts \cite{yao2023tree} and ReAct \cite{yao2023react}. On the reflection side, self-critique enables draft--revise iterations to improve robustness \cite{guo2025deepseek,jaech2024openai}. RL task further shapes these behaviors~\cite{wan2025srpo}. RLHF aligns generation with human preferences via preference-based rewards, and methods such as GRPO\cite{shao2024deepseekmath} improve optimization stability for policy learning. In reasoning-centric settings, rewards often combine final correctness with intermediate-quality signals, thereby optimizing multi-step reasoning strategies. However, most self-reflection and RL optimization operate in \emph{free-form natural language}: reflections do not map cleanly to discrete semantic-token actions, and multi-round reflection introduces high and variable inference overhead, which is undesirable for latency-sensitive retrieval.

Unlike natural-language self-reflection, GRC performs reflection and correction in the semantic-token space and optimizes these decisions with GRPO-style RL under rewards that score both final retrieval quality and the reflection--correction process.

\section{Preliminaries}
\subsection{Problem Definition}
Let $\mathcal{U}$ and $\mathcal{I}$ denote the sets of users and items, respectively. For each user $u \in \mathcal{U}$, we represent the interaction history as $\mathcal{S}_u = [i_1, i_2, \ldots, i_T]$, where $i_{t}$ is the item interacted with at step $t$ and $T$ is the maximum sequence length. Following RQ-VAE-based GR methods \cite{rajput2023recommender, yang2025sparse}, each item $i \in \mathcal{I}$ is encoded as a length $L$ semantic token sequence $\mathbf{z}_i = [z_{i,1}, z_{i,2}, \ldots, z_{i,L}]$. Here, $L$ is the number of quantization levels, and $z_{i,\ell} \in \mathcal{V}_\ell$ is a discrete token from the $\ell$-th codebook. Let $\mathcal{Z}$ be the set of valid semantic token sequences. We use a deterministic lookup function $\phi: \mathcal{Z} \rightarrow \mathcal{I}$ such that $\phi(\mathbf{z}_i) = i$. Thus, any generated sequence $\hat{\mathbf{z}} \in \mathcal{Z}$ corresponds to an item $\hat{i} = \phi(\hat{\mathbf{z}})$. Given user $u$, GR aims to generate a candidate set $\mathcal{I}_u^{\text{cand}} \subseteq \mathcal{I}$ with high coverage of the user’s future relevant items. We model a conditional distribution $p_\theta(\mathbf{z} \mid \mathcal{S}_u)$ over semantic token sequences. Given a target item $i^{\text{tar}}$ with token sequence $\mathbf{z}_{i^{\text{tar}}}$, we train by minimizing the negative log-likelihood:
\begin{equation}
\mathcal{L}_{\text{MLE}} 
= -\sum_{t=1}^{L} \log p_\theta\big(z_{i^{\text{tar}},t} \mid z_{i^{\text{tar}},<t}, \mathcal{S}_u\big),
\label{Eq:mle}
\end{equation}
where $z_{i^{\text{tar}},<t} = [z_{i^{\text{tar}},1}, \dots, z_{i^{\text{tar}},t-1}]$ and $\theta$ is the model parameters. 

\subsection{Pre-trained Generative Retrieval Backbone}
We first pre-train a standard GR model and use it as the backbone of our GRC framework.

\textbf{Model architecture.} We use an encoder–decoder Transformer \cite{rajput2023recommender}. Given $\mathcal{S}_u$, the encoder produces a user representation $\mathbf{h}_u = \mathrm{Enc}_\theta(\mathcal{S}_u)$. Conditioned on $\mathbf{h}_u$, a Transformer decoder parameterizes an autoregressive distribution over semantic token sequences,
\begin{equation}
p_\theta(\mathbf{z}\mid\mathcal{S}_u)=\prod_{t=1}^{L}p_\theta(z_t\mid z_{<t},\mathbf{h}_u),
\end{equation}
where $p_\theta(z_t\mid z_{<t},\mathbf{h}_u)$ is parameterized by $\mathrm{Dec}_\theta$. At step $t$ ($1 \le t \le L$), the decoder outputs logits over the codebook $\mathcal{V}_t$.

\textbf{Training and inference.} During pre-training, we use teacher forcing and minimize $\mathcal{L}_{\text{MLE}}$ (Eq. \ref{Eq:mle}) by conditioning each next-token prediction on the ground-truth prefix. At inference, we decode autoregressively from a special BOS token. We use beam search with beam size $B$ to generate multiple trajectories $\{\hat{\mathbf{z}}_b\}^{B}_{b=1}$ and map each to an item via $\phi(\hat{\mathbf{z}_b})$.

\textbf{Limitations.} Despite being standard, this training and inference setup has two limitations that motivate our GRC method:
\begin{enumerate}[noitemsep, topsep=0pt, leftmargin=*]
\item Teacher forcing trains the model under ground-truth prefixes, but provides no supervision for detecting or correcting deviations when decoding from its own predictions. Consequently, errors can compound at inference.
\item With multi-level semantic tokens, early mistakes can shift subsequent conditional distributions, leading to cascading errors across codebooks and reduced retrieval recall.
\end{enumerate}

These limitations motivate our GRC framework, which augments GR with SFT and RL to enable structured reflection and correction.

\begin{figure*}[htbp]
  \includegraphics[width=0.92\textwidth]{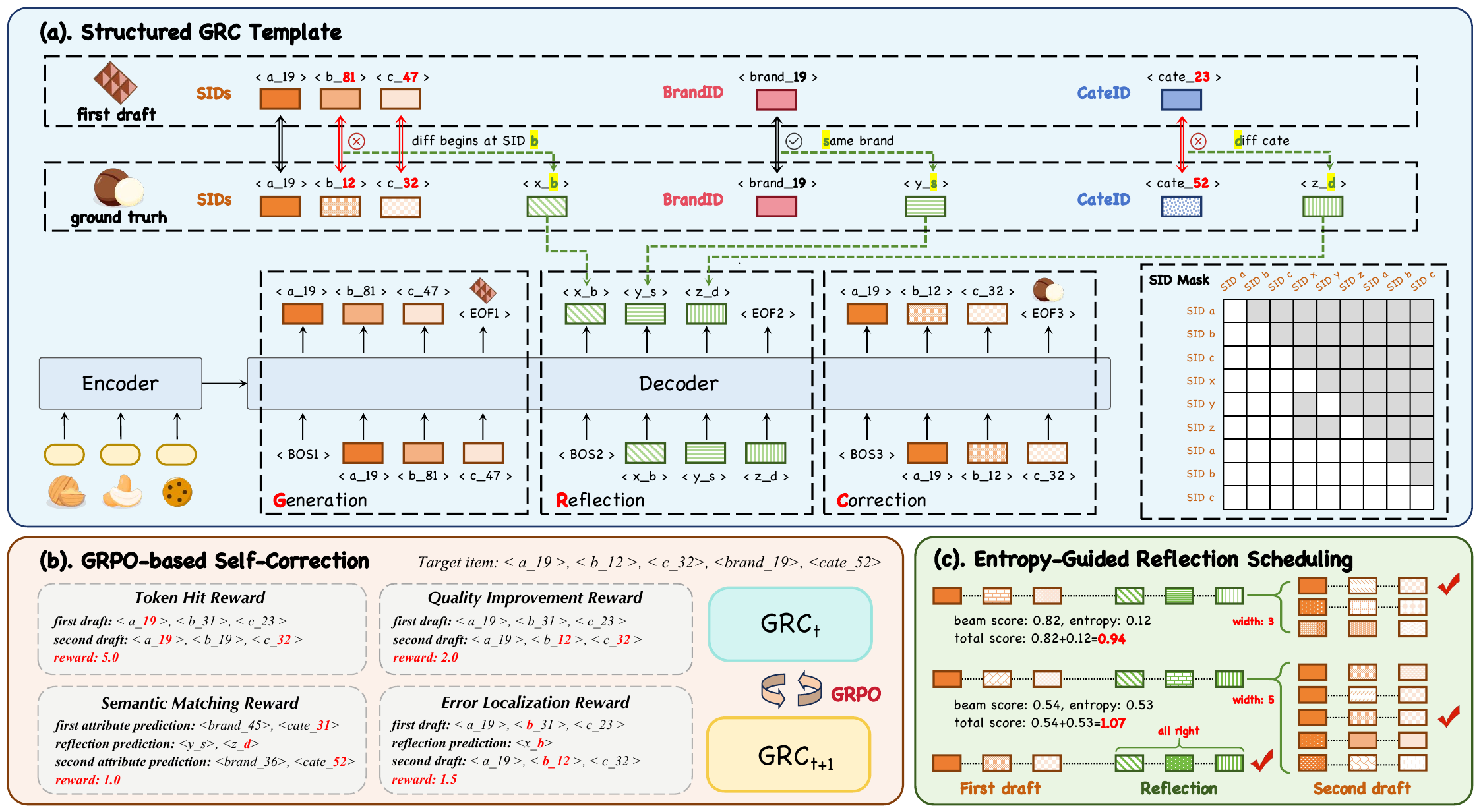}
      \vspace{-5pt}
  \caption{ 
Overview of the proposed GRC framework, which extends standard generative retrieval into a structured Generation–Reflection–Correction process with (a). Supervised reflection–correction template, (b). GRPO-based optimization of self–correction, and (c). Entropy-Guided Reflection Scheduling integrated with beam search for practical deployment.
  }
  \Description{xx.}
  \label{fig:framework}
\vspace{-10pt}
\end{figure*}
\section{Methodology}
We extend GR into a Generation–Reflection–Correction (GRC) framework to reduce error propagation in autoregressive decoding. As shown in Figure~\ref{fig:framework}, our framework consists of three components:
\begin{itemize}[noitemsep, topsep=0pt, leftmargin=*]
\item
\textbf{SFT stage (Sec.~\ref{sec:sft}).} We first introduce a \textbf{structured reflection-correction template} and perform supervised fine-tuning, formulating the reflection process as discrete decisions over token positions rather than free-form text.
\item
\textbf{RL stage (Sec.~\ref{sec:rl}).} We further optimize the model on complete generation–reflection–correction trajectories with GRPO, directly refining its self-correction policies with sequence-level rewards in the expanded trajectory space.
\item
\textbf{Inference-time scheduling (Sec.~\ref{sec:Entropy}).} We integrate GRC with beam search and allocate a fixed reflection-correction budget to low-confidence beams based on uncertainty.
\end{itemize}

\subsection{Supervised Fine-tuning with a Structured Reflection-Correction Template}
\label{sec:sft}
The GR backbone is trained with teacher forcing under ground-truth prefixes, and therefore lacks (i) the ability to identify where decoding diverges from the target and (ii) a mechanism to correct initial generations. We therefore use structured reflection signals rather than free-form text, enabling low-overhead correction that maps directly to retrieval actions.

In this section, we introduce an SFT stage with (i) multi-granular reflection labels and (ii) a structured reflection-correction template. The template factorizes decoding into \emph{generate}, \emph{reflect}, and \emph{correct} segments, making reflection a supervised, structured prediction problem. This design expands the correction decision space while keeping the number of additional decoding steps bounded.
\subsubsection{\textbf{Multi-granular reflection labels}} 
\label{sec:reflection_label}

During SFT, we employ the pre-trained backbone to generate $B$ candidate outputs for each training instance via beam search:
\begin{equation}
\{\hat{\mathbf{z}}^{(b)}\}_{b=1}^B, \quad  \hat{\mathbf{z}}^{(b)} = [\hat{z}^{(b)}_1, \ldots, \hat{z}^{(b)}_L],
\end{equation}
where $B$ is the beam size. We use these candidates to derive reflection labels. For each candidate, we annotate token-level and semantic-level reflection labels:

\paragraph{(1) Token-level Error Localization Label.} At the token level, we locate the first position where $\hat{z}^{\text{b}}$ differs from the target $\mathbf{z}^{\text{gt}}$ as
\begin{equation}
r^{(b)}_{\text{loc}} =
\min\Big(\{t\in\{1,\ldots,L\} \mid \hat{z}^{(b)}_t\neq z^{\text{gt}}_t\}\cup\{L+1\}\Big).
\end{equation}
where $r^{(b)}_{\text{loc}} =L+1$ indicates a fully correct sequence.

\paragraph{(2) Semantic-level Consistency Labels.}
To detect discrepancies between the generated item and the ground truth from a physical-attribute perspective, we introduce semantic-level consistency labels over $K$ predefined attributes with real-world meaning. Let $\hat{i}^{(b)} = \phi(\hat{\mathbf{z}}^{(b)})$ and $i^{\text{gt}} = \phi(\mathbf{z}^{\text{gt}})$ denote the candidate and ground-truth items. We consider an attribute set $\mathcal{A}$ (\textit{e.g.}, category, seller, brand). For any attribute $a_k \in \mathcal{A}$, let $f_{a_k}(i)$ map an item $i$ to its value bucket. We then define the attribute consistency label as
\begin{equation}
r^{(b)}_{\text{sem},a_k} =
    \begin{cases}
        1, & \text{if } f_{a_k}(\hat{i}^{(b)}) = f_{a_k}(i^{\text{gt}}),\\[3pt]
        0, & \text{otherwise}.
    \end{cases}
\end{equation}

Stacking these indicators produces $\mathbf{r}^{(b)}_{\text{sem}} \in \{0,1\}^{K}$. The dimensionality $K$ is extensible: additional attributes can be incorporated by defining a new $a_k$ along with its corresponding scoring function $f_{a_k}(\cdot)$. 
Finally, the structured, multi-granular reflection label for candidate $b$ is defined as follows:
\begin{equation}
\mathbf{r}^{(b)} = \big( r^{(b)}_{\text{loc}};\; \big[ r^{(b)}_{\text{sem},a_k} \big]_{k=1}^{K} \big),
\end{equation}
Here, $\big[ r^{(b)}_{\text{sem},a_k} \big]_{k=1}^{K}$ summarizes semantic-level consistency, providing structured signals for the subsequent correction module.
\subsubsection{\textbf{Structured Reflection-Correction Template}}

\label{Sec:template}

During SFT, we adopt a three-segment decoding template consisting of (i) initial prediction, (ii) multi-granular reflection, and (iii) correction, with total length $L_{\text{dec}}=2L+K+4$. We implement the template in a single decoding pass by concatenating the three segments into one sequence. Given beam candidate $\hat{\mathbf{z}}^{(b)}$, the corresponding supervised target sequence is:
\begin{small}
\begin{equation}
\begin{aligned}
\mathbf{o}^{(b)} =
    \big[
    & \hat{z}^{(b)}_1,\ldots,\hat{z}^{(b)}_L,\langle\text{EOF}_{\text{1}}\rangle,\;\; 
    r^{(b)}_{\text{loc}}, r^{(b)}_{\text{sem},a_1},\ldots,r^{(b)}_{\text{sem},a_K},\langle\text{EOF}_{\text{2}}\rangle, 
    \\& z^{\text{gt}}_1,\ldots,z^{\text{gt}}_L,\langle\text{EOF}_{\text{3}}\rangle\big],
\end{aligned}
\end{equation}
\end{small}
where the three segments correspond to initial prediction, reflection labels, and the ground-truth semantic labels, respectively, and $\langle\text{EOF}_{1}\rangle$, $\langle\text{EOF}_{2}\rangle$, and $\langle\text{EOF}_{3}\rangle$ are special tokens that delimit the three parts of the template.

\paragraph{Initial prediction segment.}
The initial prediction segment matches pre-training: triggered by $\langle\text{BOS}_{\text{1}}\rangle$, it autoregressively generates $\hat{\mathbf{z}}^{(b)}$, where positions $1,\ldots,L$ predict the $L$ semantic tokens.

\paragraph{Reflection segment.}
The reflection segment starts at position $L+2$, triggered by $\langle\text{BOS}_{\text{2}}\rangle$. With \textbf{causal masking}, the reflection tokens can attend to the entire initially predicted sequence.
In addition, we treat the $K+1$ reflection labels as conditionally independent given the shared context and predict them \textbf{in parallel} by modifying the decoder self-attention mask to block attention across reflection positions. The mask preserves cross-segment causality, and reflection tokens cannot attend to any subsequent correction or ground-truth tokens. Formally, the distribution of the $j$-th reflection label is:
\begin{equation}
p^{\text{ref}}_{\theta,j} = p_\theta\big( r^{(b)}_j \mid 
        \mathcal{S}_u,
        \hat{\mathbf{z}}^{(b)}
    \big), \quad j \in \{0,\ldots,K\}.
\end{equation}

At inference time, we predict all structured reflection tokens in parallel within a single decoder forward pass by applying the self-attention mask shown in Figure~\ref{fig:framework}, which removes intra-reflection dependencies, instead of decoding them iteratively token by token.
\paragraph{Correction segment.}
The correction segment starts at position $L+K+3$, triggered by $\langle\text{BOS}_{\text{3}}\rangle$. The correction segment generates the final sequence conditioned on $\hat{\mathbf{z}}^{(b)}$ and $\mathbf{r}^{(b)}$, and is trained with standard teacher forcing.

Although the correction segment still uses an autoregressive Transformer, it differs from pre-training by conditioning on two additional signals: (i) initial candidate sequence $\hat{\mathbf{z}}^{(b)}$, (ii) multi-dimensional reflection labels $\mathbf{r}^{(b)}$. The conditional distribution is:
\begin{equation}
p^{\text{cor}}_{\theta,t}
= p_\theta\!\left(
z^{\text{gt}}_{t}\mid 
\mathcal{S}_u,
z^{\text{gt}}_{<t},
\hat{\mathbf{z}}^{(b)}, \mathbf{r}^{(b)}
\right).
\end{equation}

This makes the correction distribution explicitly depend on the model’s initial generation and reflection signals, rather than only the user history.
The three-segment template allows us to supervise generation, reflection, and correction in a single pass. We optimize the SFT objective:
\begin{equation}
\begin{aligned}
\mathcal{L}_{\text{SFT}}
= &\mathcal{L}_{\text{MLE}}  + \lambda_{\text{rc}} \mathcal{L}_{\text{rc}},\\[3pt]
\mathcal{L}_{\text{rc}}
= - \sum_{j=0}^{K}& \log p^{\text{ref}}_{\theta,j}
   - \sum_{t=1}^{L} \log p^{\text{cor}}_{\theta,t}.
\end{aligned}
\label{Eq.sft}
\end{equation}

In Eq.\ref{Eq.sft}, we apply $\mathcal{L}_{\text{MLE}}$ to the initial prediction segment and use $\mathcal{L}_{\text{rc}}$ to supervise reflection and correction. The coefficient $\lambda_{\text{rc}}$ controls the overall weight of this additional supervision.

Overall, SFT trains the model to follow a structured generate, reflect, and correct procedure, which provides a strong initialization for subsequent RL.

\subsection{GRPO-Enhanced Exploration of GRC Correction Space}
\label{sec:rl}
In Section \ref{sec:sft}, we presented an SFT stage with a structured generate-reflect-correct template that provides basic self-correction behavior. This supervision is driven by offline-constructed labels and does not provide an end-to-end signal for the success of the full generate-reflect-correct trajectory, nor does it encourage exploration of alternative correction strategies.

To address these issues, we apply GRPO to optimize the model on full generate-reflect-correct trajectories. We keep the GRPO algorithm unchanged, and instead tailor the episode definition and rewards to encourage a diagnose-then-correct behavior.

\subsubsection{\textbf{Episode Definition for GRC}}
We treat the entire GRC procedure, from $\mathcal{S}_u$ to the final corrected output, as a single RL episode. Given $\mathcal{S}_u$ and the ground-truth sequence $\mathbf{z}^{\text{gt}}$, each episode consists of three steps. First, conditioned on $\mathcal{S}_u$ and initialized with $\langle\text{BOS}_{1}\rangle$, the policy samples an initial sequence $\hat{\mathbf{z}}^{(0)} \sim \pi_\theta(\mathbf{z} \mid \mathcal{S}_u)$. Second, conditioned on this initial draft $\hat{\mathbf{z}}^{(0)}$, the policy generates structured reflection tokens that characterize where and how the trajectory deviates from the target:
\begin{equation}
\hat{\mathbf{r}}
= \big(\hat{r}_{\text{loc}};\, [\hat{r}_{\text{sem},a_k}]_{k=1}^K\big)
\sim \pi_\theta(\mathbf{r} \mid \hat{\mathbf{z}}^{(0)}, \mathcal{S}_u).
\end{equation}

The ground-truth reflection label $\mathbf{r}^{\text{gt}}$ (Sec.~\ref{sec:reflection_label})  is then used to compute the reflection-oriented components of the reward. Finally, initialized with $\langle\text{BOS}_{\text{3}}\rangle$, the policy samples a corrected sequence $\hat{\mathbf{z}}^{(1)} \sim \pi_\theta\big(\mathbf{z} \mid \hat{\mathbf{z}}^{(0)}, \hat{\mathbf{r}}, \mathcal{S}_u \big)$. We denote the entire trajectory as
\begin{equation}
\tau = \big(\mathcal{S}_u,\, \hat{\mathbf{z}}^{(0)},\, \hat{\mathbf{r}},\, \hat{\mathbf{z}}^{(1)}\big).
\end{equation}

In this way, GRPO optimizes the reflection–correction policy in the enlarged trajectory space, rather than merely imitating the supervised correction template.
\subsubsection{\textbf{Reward Shaping for the GRC Trajectory}}
In this part, we design rewards to (i) decouple generation quality from correction behavior and (ii) leverage the reflection labels (Sec \ref{sec:reflection_label}) to evaluate both the final output and the intermediate reflection and correction decisions. Formally, the total reward is defined as:
\begin{equation}
R = R_{\text{task}} + \beta_{\text{cor}} R_{\text{cor}},
\end{equation}
where $R_{\text{task}}$ evaluates generation quality, $R_{\text{cor}}$ evaluates reflection and correction quality, and $\beta_{\text{cor}}$ controls their relative importance.

\smallskip
\noindent\textbf{(i) Task Reward: Token-Level Retrieval Gain.}

Due to autoregressive dependencies, early errors often propagate to later positions and causes error accumulation \cite{rajput2023recommender}. Motivated by this, we use a token-level hit count as the task signal, computed separately for the initial and corrected sequences as:
\begin{equation}
\begin{aligned}
\ell_0 = \sum_{t=1}^{L} \mathbb{I}\!\left[\hat{z}^{(0)}_{t} = z^{\text{gt}}_{t}\right], 
\ell_1 = \sum_{t=1}^{L} \mathbb{I}\!\left[\hat{z}^{(1)}_{t} = z^{\text{gt}}_{t}\right],
\end{aligned}
\end{equation}
We then define the task reward as follows, placing greater emphasis on the corrected sequence via the coefficient $\beta_{\text{last}}$.
\begin{equation}
R_{\text{task}} = \ell_0 + \beta_{last} \ell_1.
\end{equation}

\smallskip
\noindent\textbf{(ii) Correction Reward: Trajectory-Level Reflection Gain.}

To further shape correction behavior, we leverage the reflection labels (Sec.~\ref{sec:reflection_label}) and the difference between $\hat{z}^{(0)}_t$ and $\hat{z}^{(1)}_t$ to evaluate correction behavior along three dimensions:
\begin{equation}
R_{\text{cor}} = \beta_{\text{loc}} R_{\text{loc}} + \beta_{\text{sem}} R_{\text{sem}} + R_{\Delta},
\end{equation}
where $\beta_{\text{loc}}$ and $\beta_{\text{sem}}$ weight the $R_{\text{loc}}$ and $R_{\text{sem}}$, respectively.

\emph{\textbf{(1) Error Localization Reward $R_{\text{loc}}$.}}
This component rewards both accurate error localization and effective correction within the predicted region. Let $\hat{r}_{\text{loc}}$ denote the predicted first-divergence position, and $r^{\text{gt}}_{\text{loc}}$ denote the real first-divergence position. We define the label-accuracy reward as follows:
\begin{equation}
R_{\text{loc-label}} = \mathbb{I}[\hat{r}_{\text{loc}} = r^{\text{gt}}_{\text{loc}}],
\end{equation}
where $\mathbb{I}[\cdot]$ is the indicator function. We define the predicted error region $\mathcal{E}^{\text{loc}}_{\text{pred}}= \{t \mid t \ge \hat{r}_{\text{loc}}\}$ and the set of actually corrected positions as $\mathcal{C}^{\text{fix-loc}} = \big\{ t \mid \hat{z}^{(0)}_t \neq z^{\text{gt}}_t,\ \hat{z}^{(1)}_t = z^{\text{gt}}_t \big\}$. We then quantify position-level correction using the hit ratio $R_{\text{pos-cor}}$:
\begin{equation}
R_{\text{loc-cor}}
   = \frac{|\mathcal{E}^{\text{loc}}_{\text{pred}} \cap \mathcal{C}^{\text{fix-loc}}|}
          {|\mathcal{E}^{\text{loc}}_{\text{pred}}| + \varepsilon},
\end{equation}
where $\varepsilon>0$ is a small constant. Accordingly, we define the error localization reward as $R_{\text{loc}} = R_{\text{loc-label}} + R_{\text{loc-cor}}$.

\emph{\textbf{(2) Semantic Matching Reward $R_{\text{sem}}$.}}
This term rewards accurate semantic reflection and subsequent semantic alignment after correction. The semantic-label accuracy is
\begin{equation}
R_{\text{sem-label}}
   = \frac{1}{K} \sum_{k=1}^{K}
     \mathbb{I}\big[\hat{r}_{\text{sem},a_k} = r^{\text{gt}}_{\text{sem},a_k}\big].
\end{equation}

We further reward attributes that are initially predicted as mismatched (i.e. $\hat{r}_{\text{sem},a_k}=0$) but become matched after correction. Let $\hat{i}^{(0)}=\phi(\hat{\mathbf{z}}^{(0)})$, $\hat{i}^{(1)}=\phi(\hat{\mathbf{z}}^{(1)})$, and $i^{\text{gt}}$ denote the initial prediction, the corrected prediction, and the target item, respectively. For each attribute $a_k$, we define a semantic-level correction indicator $R_{\text{sem-cor},a_k}$ and average it over attributes to obtain $R_{\text{sem-cor}}$:
\begin{equation}
\begin{aligned}
&R_{\text{sem-cor}}
   = \frac{1}{K} \sum_{k=1}^{K} R_{\text{sem-cor},a_k}, \\
R_{\text{sem-cor},a_k}
   &= \mathbb{I}\big[
        \hat{r}_{\text{sem},a_k} = 0\ \land\ 
        f_{a_k}(\hat{i}^{(0)}) \neq f_{a_k}(i^{\text{gt}})\ \land\ \\
       &  f_{a_k}(\hat{i}^{(1)}) = f_{a_k}(i^{\text{gt}})
     \big], \\
\end{aligned}
\end{equation}
where $\land$ denotes logical conjunction. The semantic matching reward is set as $R_{\text{sem}} = R_{\text{sem-label}} +R_{\text{sem-cor}}$.

\emph{\textbf{(3) Quality Improvement Reward $R_{\Delta}$.}}
We further reward strict improvement and define $R_{\Delta}$ as
\begin{equation}
R_{\Delta} =\begin{cases}
   \ell_1 - \ell_0, & \text{if } \ell_1 > \ell_0, \\
   0,               & \text{otherwise}.
   \end{cases}
\end{equation}


Together, these rewards provide trajectory-level signals over both generation quality and correction behavior, allowing GRPO to effectively optimize the reflection–correction policy.

\subsubsection{\textbf{Training Pipeline}}
Given the above episode and reward definitions, we train GRC with GRPO using advantages computed within each user group. The training proceeds in three steps:

\textbf{(1) Initialization and rollout.} We initialize the policy with the SFT weights, i.e., $\theta_0 \leftarrow \theta_{\text{SFT}}$. At each iteration, we sample a minibatch of user histories and roll out one or more trajectories per user, obtaining episode $\{\tau_i\}$ and their rewards $\{R_i\}$.

\textbf{(2) Group-wise advantage estimation.} We group episodes by user and compute a group-relative advantage $A_i$ by normalizing and ranking $\{R_j\}$ within each group, which reduces sensitivity to reward scale and stabilizes optimization. $A_i$ is define as:
\begin{equation}
A_i = \mathrm{GRAdv}\big(R_i;\{R_j\}_{j\in\text{group}(i)}\big),
\end{equation}
where $\mathrm{GRAdv}(\cdot)$ denotes the GRPO group-relative advantage computed by rank-normalizing returns within each user group.

\textbf{(3) GRPO objective.} We then update the policy with the GRPO clipped objective and a standard KL penalty:

\begin{equation}
\begin{aligned}
\mathcal{L}_{\text{GRPO}}(\theta)
   &= \mathbb{E}_i \Big[
       \min\big(
         \frac{\pi_\theta}{\pi_{\theta_{\text{old}}}} A_i,\,
         \mathrm{clip}\big(\frac{\pi_\theta}{\pi_{\theta_{\text{old}}}}, 1-\epsilon, 1+\epsilon\big) A_i
       \big) \\
      &- \beta_{kl}\, \mathrm{KL}\big(\pi_\theta\ \|\ \pi_{\theta_{\text{ref}}}\big)
     \Big]
\end{aligned}
\end{equation}
where $\pi_{\theta_{\text{old}}}$ denotes the old policy, $\pi_{\theta_{\text{ref}}}$ is the reference policy, $\pi_{\theta}$ is the current policy and $\beta_{kl}$ controls the strength of the KL penalty.

\subsection{Entropy-Guided Reflection Scheduling}
\label{sec:Entropy}
In large-scale recommender systems, GR is typically deployed with beam
search to balance effectiveness and latency. However, running a full
reflection–correction procedure on every beam is computationally
prohibitive. We therefore propose an entropy-guided reflection
scheduling (EGRS) strategy that is tightly integrated with beam search
and allocates limited reflection capacity to trajectories that are more
likely to benefit from self-correction.

At inference time, for each user $u$, we first apply beam search to
generate $B$ initial candidate paths
$\{\hat{\mathbf{z}}^{(0)}_b\}_{b=1}^{B}$, where
$\hat{\mathbf{z}}^{(0)}_b$ denotes the semantic token sequence of the
$b$-th beam with length $L$. We then enter the reflection stage, where
the model generates reflection tokens for each path and leverages their
uncertainty to guide subsequent beam selection. \textbf{For paths whose
reflection explicitly predicts that the initial generation is already
correct, we skip further regeneration and retain the original
trajectory.}

For the $b$-th beam, let
$\hat{\mathbf{r}}_b = (r_{b,1}, \dots, r_{b,T_r})$ denote the generated
reflection token sequence. We quantify reflection uncertainty using the average token-level entropy during reflection: 
\begin{equation}
\label{eq:avg_ref_entropy}
\bar{H}^{\text{ref}}_b
= \frac{1}{T_r}\sum_{t=1}^{T_r}
\left(
- \sum_{v \in \mathcal{V}^{\text{ref}}_t}
p_\theta\!\left(v \mid \hat{\mathbf{z}}^{(0)}_b, \mathcal{S}_u, t\right)
\log p_\theta\!\left(v \mid \hat{\mathbf{z}}^{(0)}_b, \mathcal{S}_u, t\right)
\right),
\end{equation}
where $T_r = K+1$ is the number of reflection slots, and $t \in \{1,\ldots,T_r\}$ indexes the
slot. $\mathcal{V}^{\text{ref}}_t$ denotes the vocabulary for slot $t$. A larger $\bar{H}^{\text{ref}}_b$ indicates lower confidence in the
reflection tokens for path $b$, suggesting that its current correction
state is more uncertain and potentially more error-prone.


We use the average reflection entropy $\bar{H}^{\text{ref}}_b$ to re-rank beams during correction. For each beam path $b$, let $\text{score}^{\text{base}}_b$ be the standard beam score, i.e., the sum of log-probabilities along its draft trajectory $\hat{\mathbf{z}}^{(0)}_b$. We define an entropy-calibrated score:
\begin{equation}
\label{eq:egrs_score}
\text{score}^{\text{EGRS}}_b
= \text{score}^{\text{base}}_b + \alpha_{\text{e}} \cdot \bar{H}^{\text{ref}}_b,
\end{equation}
where $\alpha_{\text{e}}>0$ controls the strength of entropy calibration. During correction, we maintain a fixed beam budget $B$ and prune candidates by
\(
B \leftarrow \mathrm{Top}\text{-}B\big(\{\text{score}^{\text{EGRS}}_b\}_{b\in B}\big)
\)
instead of $\text{score}^{\text{base}}_b$. As a result, high-entropy paths are more likely to survive pruning and thus receive more downstream correction computation under the same beam budget, while low-entropy paths are discarded earlier. This yields an entropy-aware scheduling mechanism without changing the standard beam-search interface, making it easy to tune under practical latency constraints.
\section{Offline Experiments}
In this section, we conduct extensive experiments on three datasets to evaluate the effectiveness of the proposed method and to answer the following research questions:
\begin{itemize}[noitemsep, topsep=0pt, leftmargin=*]
\item \textbf{RQ1}: How does GRC perform in comparison with other state-of-the-art  retrieval models?
\item \textbf{RQ2}: How does each component contribute to the overall performance of GRC?
\item \textbf{RQ3}: How do hyperparameters influence the model performance?
\item \textbf{RQ4}: How well does GRC perform in online environments?
\end{itemize}

\begin{table*}[t]
\centering
\caption{Performance comparison across different recommendation paradigms. $GC_{SFT}$ denotes the single two-pass decoding variant without reflection labels, $GRC_{SFT}$ denotes our method after supervised fine-tuning with the structured template, and $GRC_{RL}$ further leverages GRPO to explore and refine self-correction strategies. The best metric for each dataset is highlighted in bold, and the second-best is underlined. The last row reports the relative improvement of $GRC_{RL}$ over the strongest baseline.}
\vspace{-5pt}
\renewcommand{\arraystretch}{0.90} 
\resizebox{0.98\linewidth}{!}{
    \begin{tabular}{lcccccccccccc}
    \toprule
    \multirow{2}{*}{\textbf{Method}} & \multicolumn{4}{c}{\textbf{Arts}} & \multicolumn{4}{c}{\textbf{Instruments}} & \multicolumn{4}{c}{\textbf{Industrial Dataset}}\\
\cmidrule(lr){2-5} \cmidrule(lr){6-9} \cmidrule(lr){10-13}
     & R@5 & N@5 & R@10 & N@10 &  R@5 & N@5 & R@10 & N@10 & R@5 & N@5 & R@10 & N@10  \\ 
    \midrule \midrule
    DSSM &0.0611 &0.0436 &0.0833 &0.0461 &0.0618 &0.0447 &0.0845 &0.0528 &0.0954&0.0689&0.1204&0.0913\\ 
    SASRec &0.0668 &0.0483 &0.0906 &0.0523 &0.0690 &0.0583 &0.0918 &0.0657 &0.0997&0.0726&0.1288&0.0942\\ 
    HSTU &0.0761 &0.0558 &0.1027 &0.0616 &0.0804 &0.0636 &0.1023 &0.0702 &0.0895&0.0654&0.1152&0.0873 \\
    TIGER &0.0796 &0.0577&0.1051 &0.0634 &0.0849&\underline{0.0715}&0.1078&\underline{0.0776} &\underline{0.1131}&\underline{0.0828}&0.1383&0.0997 \\ 
    ReaRec &0.0762&0.0541&0.1037 &0.0612 &0.0825 & 0.0678 &0.1064 &0.0747 &0.1055&0.0779&0.1322&0.0975\\ 
    COBRA &\underline{0.0812}&\underline{0.0612} &\underline{0.1095}&\underline{0.0682} &\underline{0.0861}&0.0688&\underline{0.1095}&0.0763 &0.1090&0.0802&\underline{0.1438}&\underline{0.1063}\\ 
    \midrule
    \textbf{$GC_{SFT}$}&$0.0811$&$0.0597$ &$0.1074$ &$0.0661$&$0.0875$&$0.0743$ &$0.1102$ &$0.0791$ &$0.1164$ &$0.0837$  &$0.1409$ &$0.1039$ \\ 
    \textbf{$GRC_{SFT}$}&0.0832&0.0628 &0.1132 &0.0715 &0.0912 &0.0771 &0.1148&0.0812 &0.1207&0.0856&0.1512&0.1074 \\ $\textbf{GRC}_{RL}$&\textbf{0.0868}&\textbf{0.0667}&\textbf{0.1191}&\textbf{0.0749} &\textbf{0.0946}&\textbf{0.0794}&\textbf{0.1208}&\textbf{0.0873} &\textbf{0.1309}&\textbf{0.0935}&\textbf{0.1648}&\textbf{0.1165} \\ 
    \midrule
    Improv. &+6.90\%&+8.99\%&+8.77\%&+9.82\% &+9.87\%&+11.05\%&+10.32\%&+12.50\% &+15.74\%&+12.92\%&+14.60\%&+9.60\% \\
    \bottomrule
    \end{tabular}
}
\vspace{-3pt}
\label{table:offline_experiment_results}
\end{table*}

\begin{table}[htbp]
    \centering
    \captionsetup{skip=5pt}
    \caption{Reflection Quality on the Industrial Dataset.}
    \renewcommand{\arraystretch}{1}
    \label{table:ref_quality}
    \resizebox{230pt}{!}{  
        \begin{tabular}{c|c|c|c|c}
            \toprule
            Method &             $\mathrm{Acc}_{\mathrm{loc}}@5$ &
            $\mathrm{Acc}_{\mathrm{loc}}@10$ &
            $\mathrm{Acc}_{\mathrm{cat}}@5$ &
            $\mathrm{Acc}_{\mathrm{cat}}@10$ \\
            \midrule
            $GRC_{SFT}$ & 56.32\% & 53.35\% & 74.42\% & 71.85\%  \\
            $GRC_{RL}$ & 61.26\% & 59.94\% & 77.73\% & 76.20\%  \\
            \bottomrule
        \end{tabular}
    }
    \vspace{-19pt}
\end{table}

\subsection{Experimental Setup}
\label{sec:exp}
\textbf{Dataset.} We evaluate our proposed method on three datasets. The first two are publicly available benchmark datasets, and the third is an in-house industrial dataset.
\begin{itemize}[noitemsep, topsep=0pt, leftmargin=*]
\item \textbf{Amazon Product Reviews datasets.} To evaluate our method, we use public benchmarks derived from the Amazon Product Reviews dataset~\cite{AmazonDataset}, which contains user reviews and item metadata from May 1996 to July 2014. We focus on two categories: Arts and Musical Instruments. Arts includes 390 thousand interactions from 45 thousand users over 21 thousand items, while Musical Instruments contains 206 thousand interactions from 25 thousand users over 10 thousand items. Following the standard leave-one-out protocol~\cite{kang2018self}, we use the last item of each sequence for testing, the second-to-last for validation, and the remaining interactions for training. 

\item \textbf{Industrial Dataset.} We construct an in-house offline dataset from sequential user behaviors and feedback logs on a large-scale e-commerce advertising platform. It covers multiple traffic scenarios (\textit{e.g.}, homepage feeds, product detail pages, and interactive gaming sections) and spans March–August 2025. The dataset contains over 1 billion interactions from 19 million users over 25 million advertisements, providing broad coverage of real-world recommendation behavior. 
\end{itemize}


\noindent \textbf{Baselines.}
To evaluate the effectiveness of our method, we compare GRC with \textbf{six representative baselines}, summarized below.
\begin{itemize}[noitemsep, topsep=0pt, leftmargin=*]
\item \textbf{DSSM \cite{10.1145/2505515.2505665}}\ : A dual-tower model that augments user and item embeddings with side information features and incorporates user behavior sequence modeling to capture richer user preferences.
\item \textbf{SASRec \cite{kang2018self}}\ : A Transformer-based model that employs the self-attention mechanism to capture long-term dependencies.
\item \textbf{HSTU \cite{10.5555/3692070.3694484}}\ : A generative retrieval model that first flattens all ID-type features on the input side and directly generates item IDs from dense user representations, without constructing semantic IDs. In contrast, our method still relies on semantic ID compression and focuses on self-correction in the token space.
\item \textbf{TIGER \cite{rajput2023recommender}}\ : A GR model with the encoder– decoder architecture that constructs semantic IDs via RQ-VAE, and employs an autoregressive decoder to generate items.
\item \textbf{ReaRec \cite{rearec}}\ : A transformer-based model that generate implicit intermediate outcomes without explicitly defined inference paths.
\item \textbf{COBRA \cite{yang2025sparse}}\ : A generative retrieval model with a decoder-only architecture, which additionally leverages dense user embeddings for nearest-neighbor retrieval to expand the candidate set when the generative model returns an insufficient number of items.
\item \textbf{Ours / Ablations}\ :$GRC_{SFT}$ removes RL from $GRC_{RL}$, and $GC_{SFT}$ further removes the reflection labels from $GRC_{SFT}$ while keeping the same two-pass draft-and-correct prediction procedure.
\end{itemize}

\noindent \textbf{Evaluation Metrics.}
For evaluation, we adopt  \(Recall@K (R@K)\) and \(NDCG@K (N@K) \) as standard recommendation metrics, with $K$ set to 5 and 10. To align our offline evaluation with the model’s online performance in real-world industrial scenarios, we additionally report \(Recall@100\) and \(Recall@200\) in the ablation study section. 


\noindent\textbf{Implementation Details.}
This section describes the hyperparameter settings in our experiments. The training process is conducted on a distributed Pytorch \cite{pytorch} platform with 2 parameter servers and 10 workers, each equipped with a single Nvidia A100 GPU. To ensure a fair comparison, we keep the total number of Transformer layers and the hidden size consistent across generative models. Semantic attribute features are included in the encoder input for all compared methods, so that \textbf{performance gains of GRC will not be attributed to additional attribute information.} We employ the pretrained Qwen3-8B model \cite{qwen3} to encode item embeddings and train an RQ-VAE with a 4-layer codebook (300 codewords per layer) on each dataset. The codebook is fixed after training, and all GR methods share it to obtain semantic IDs for both training and inference. Encoder–decoder models are configured with one encoder layer and two decoder layers, while COBRA, as a decoder-only model, uses three stacked decoder layers. For all experiments, we use an embedding size of 128 and a hidden size of 640 for fair comparison. In the SFT stage, we set the weight of $\mathcal{L}_{\text{rc}}$ in the loss function to \(\lambda_{rc}=1.2\). In the RL stage, we set the clipping parameter $\epsilon$ to 0.15 and the regularization coefficient $\beta$ to 0.03, which provide a reasonable trade-off between exploration and exploitation. Since policy optimization is not the main focus of this work, we do not conduct a dedicated sensitivity study for these two parameters. In reward design, we set \(\beta_{cor}\) to 2.2, placing more emphasis on reflection and correction quality than on the initial generation quality. We further introduce a hyperparameter \(\beta_{last}\) to balance the contributions of the final and initial generations in the reward. Based on empirical observations that up-weighting the final generation improves performance, we set \(\beta_{last}=2.0\). Then, we introduce \(\beta_{loc}\) and \(\beta_{sem}\), which control the position-level and semantic-level correction rewards, respectively. We select their values via grid search and use \(\beta_{loc}=1.0\) and \(\beta_{sem}=0.8\). At inference time, we use beam search with \(B=200\) for both offline evaluation and online serving. The entropy coefficient \(\alpha_{e}\) is set to 0.2. Regarding the attribute set used for semantic-level consistency, we use the leaf category (no extra annotation) on the public datasets, and use both the leaf category and brand on the industrial dataset.

\subsection{Experiment Performance (RQ1)}
Table \ref{table:offline_experiment_results} reports the overall performance of GRC against multiple baselines on both public and industrial datasets. The best results are highlighted in bold, and the second-best results are underlined. Based on these results, we make the following observations:
\begin{itemize}[noitemsep, topsep=0pt, leftmargin=*]
\item \textbf{Overall effectiveness.} $\mathbf{GRC}_{RL}$ consistently outperforms all baselines on three datasets, with \emph{average relative gains} of \textbf{10.84\%} (R@5), \textbf{10.99\%} (N@5), \textbf{11.23\%} (R@10), and \textbf{10.64\%} (N@10) over the strongest baseline. The improvements are most pronounced on the Industrial dataset, where GRC exceeds \textbf{15\%} in R@5, highlighting its practical value for large-scale recommendation. The results further highlight the advantage of explicit reflection and correction in reducing error propagation during decoding.
\item \textbf{Generative retrieval vs. Representation-based retrieval.} 

TIGER and COBRA consistently outperform DSSM and SASRec, confirming the advantage of generative retrieval in large-scale recommendation. Among generative baselines, COBRA attains the strongest performance on most datasets by leveraging approximate nearest neighbor search to enrich the candidate set, thereby enhancing generalization. In contrast, HSTU degrades sharply on the industrial dataset, suggesting that directly generating raw item IDs is inadequate when the item corpus is extremely large (over $10^3$ times larger than the public datasets). This observation highlights the necessity of semantic vocabulary compression for generative retrieval in real-world settings.
\item \textbf{Impact of Reflection labels in GRC.} We verify the role of reflection with two observations. First, $GRC_{\text{SFT}}$ consistently outperforms $GC_{\text{SFT}}$ across all datasets, showing that explicit, structured reflection improves the subsequent correction beyond what can be achieved by an extra decoding pass. Second, we measure reflection quality by the average accuracy over the top-$K$ first-pass drafts ($K\!\in\!\{5,10\}$), as shown in Table~\ref{table:ref_quality}. $Acc_{\text{loc}}@K$ for first-divergence localization and $Acc_{\text{cat}}@K$ for category-consistency prediction. $GRC_{\text{RL}}$ yields higher $Acc_{\text{loc}}@K$ and $Acc_{\text{cat}}@K$ than $GRC_{\text{SFT}}$, indicating that RL improves reflection quality, which is consistent with its better final recommendation performance.
\item \textbf{Impact of SFT and RL stage in GRC.} Our GRC surpasses TIGER and COBRA on all datasets and metrics, showing the benefit of explicitly modeling a Generate–Reflect–Correct process. \textbf{Since $\mathbf{GRC}_{SFT}$ is implemented by augmenting TIGER with our structured reflection–correction template, TIGER serves as a natural baseline of $\mathbf{GRC}_{SFT}$}. The results show that $GRC_{SFT}$ yields consistent gains over TIGER across all datasets and metrics, and matches or exceeds COBRA in most settings (with a minor gap on R@5 for Arts). $GRC_{RL}$ further optimizes the full GRC trajectory via GRPO using both token-level and trajectory-level rewards, and achieves state-of-the-art performance across all datasets and metrics.
\end{itemize}

\begin{table}[htbp!]
\caption{Ablation Study of GRC.}
\vspace{-5pt}
\renewcommand{\arraystretch}{0.99}
\centering
\resizebox{200pt}{!}{%
\begin{tabular}{l|cccc}
\toprule
Method & R@5 & R@10 & R@100 & R@200 \\
\midrule
\textbf{$GRC_{RL}$} & \textbf{0.1309} & \textbf{0.1648} & \textbf{0.2928} & \textbf{0.3576}\\
$GRC_{SFT}$ & 0.1207 & 0.1512 & 0.2735 & 0.3389 \\
w/o $R_{task}$ & 0.0973 & 0.1126 & 0.2341 & 0.2953\\
w/o $R_{cor}$ & 0.1238 & 0.1573 & 0.2776 & 0.3412 \\
w/o $R_{loc}$ in $R_{cor}$ & 0.1275 & 0.1621 & 0.2873 & 0.3519 \\
w/o $R_{sem}$ in $R_{cor}$ & 0.1282 & 0.1614 & 0.2864 & 0.3497 \\
w/o $R_{\Delta}$ in $R_{cor}$ & 0.1244 & 0.1598 & 0.2827 & 0.3463 \\
w/o $EGRS$ & 0.1287 & 0.1629 & 0.2801 & 0.3428 \\
\bottomrule
\end{tabular}%
}
\label{table:ablation_experiment_results}
\end{table}
\subsection{Ablation Study (RQ2)}
To investigate the effectiveness of each component in the proposed model, we conduct a series of ablation studies on the industrial dataset, which represents the most complex and representative scenario on our recommendation platform. As shown in Section \ref{sec:exp}, the industrial dataset is orders of magnitude larger in scale than public datasets, providing a more challenging evaluation environment.

\begin{itemize}[noitemsep, topsep=0pt, leftmargin=*]
\item \textbf{w/o $R_{task}$}, remove the task reward so that RL focuses solely on encouraging the reflection–correction ability.
\item \textbf{w/o $R_{cor}$}, remove the correction reward so that RL only optimizes the generation quality of the entire trajectory.
\item \textbf{w/o $R_{loc}$}, drops the error localization term, preventing the model from being rewarded for localizing the first erroneous token.
\item \textbf{w/o $R_{sem}$}, removes the semantic matching term, ignoring whether the model correctly identifies and fixes semantic mismatches across bucketed attributes (\textit{e.g.}, category, brand).

\item \textbf{w/o $R_{\Delta}$}, discards the quality improvement term, so the reward no longer enforces strict improvement of the final sequence and cannot distinguish trajectories that merely detect errors from those that actually produce a better corrected sequence.
\item \textbf{w/o} $EGRS$, keeps all inference settings identical to the full model, but disables EGRS by setting $\alpha_e=0$ in Eq.~\ref{eq:egrs_score} (i.e., removing the entropy term). As a result, beam pruning relies solely on the standard beam-search score, without using reflection entropy.
\end{itemize}

Table~\ref{table:ablation_experiment_results} yields the following observations:
\begin{itemize}[noitemsep, topsep=0pt, leftmargin=*]
\item Overall, removing any module leads to a noticeable degradation, indicating that each part of our design contributes meaningfully to the final performance.
\item If we remove the task reward $R_{task}$, the model suffers a severe performance collapse. In this case, the RL objective is no longer anchored to the original next-item prediction goal and optimizes only reflection–correction signals, which we empirically observe to induce reward hacking behaviors and an unstable policy. This indicates that \textbf{$R_{task}$ is indispensable for keeping the RL optimization aligned with the underlying retrieval objective}. Among the remaining terms, discarding the semantic-correction reward $R_{cor}$ causes the largest drop across all metrics, highlighting that explicitly rewarding successful reflection and correction is crucial for effective RL optimization. Within $R_{cor}$, removing the quality-improvement subterm $R_{\Delta}$ yields the most pronounced decline, confirming that encouraging positive pre–post quality gain is the key driver of these improvements.
\item Disabling the EGRS substantially degrades R@100 and R@200, showing that dynamically allocating reflection budget to high-uncertainty decoding trajectories effectively directs limited computation to promising corrections, yielding \textbf{higher reflection success rate} under a fixed budget.
\end{itemize}

\vspace{-2pt}
\begin{figure}[htbp]
  \includegraphics[width=0.44\textwidth]{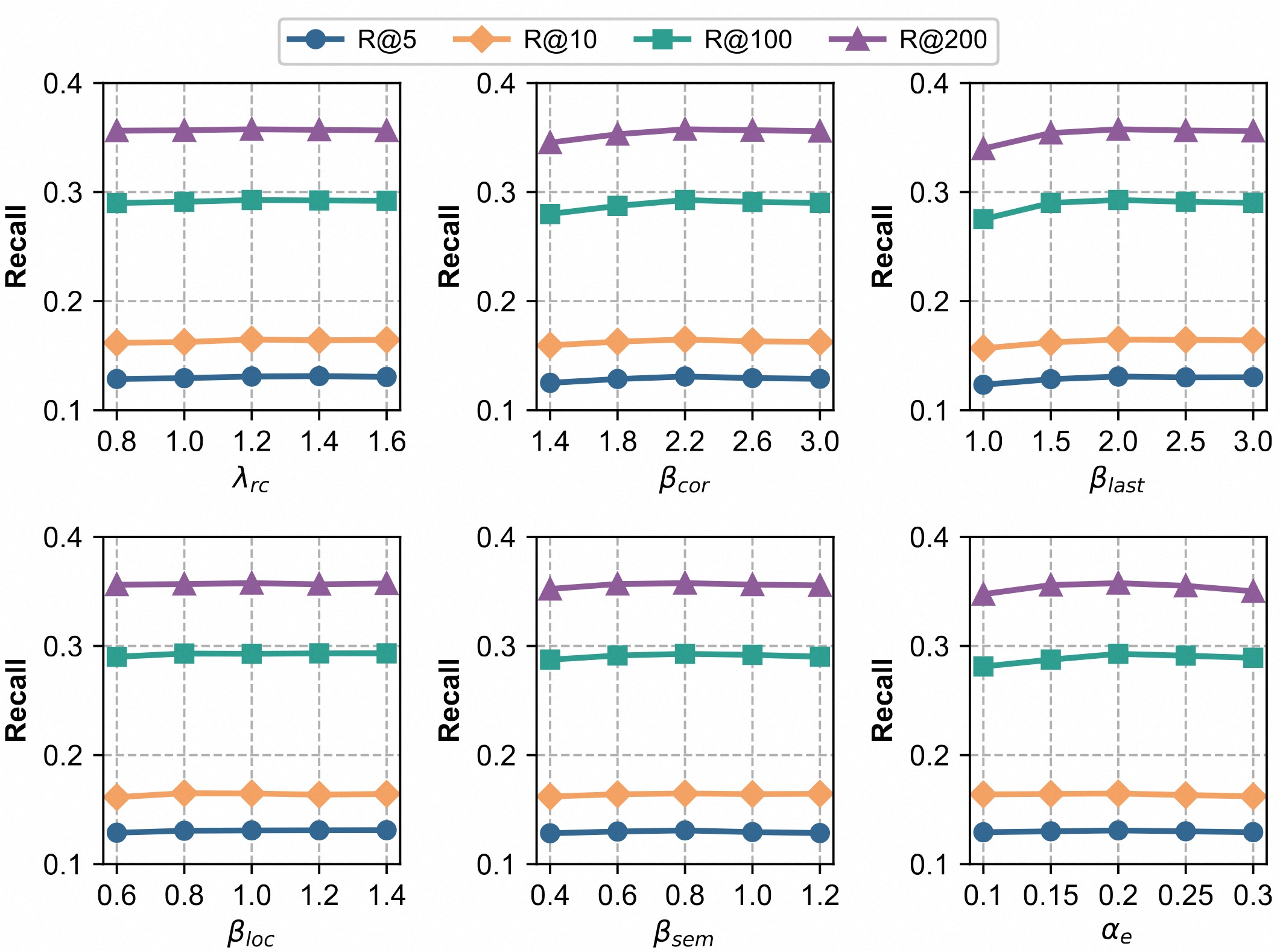}
  \vspace{-8pt}
  \caption{Sensitivity Analysis}
  \Description{xx.}
  \label{fig:sensitive_v1}
\end{figure}
\vspace{-15pt}
\subsection{Sensitivity Analysis (RQ3)}
~\label{sec:Sensitivity}
In this section, we analyze the sensitivity of our model to several key hyperparameters. Specifically, we focus on the coefficient \(\lambda_{rc}\) in the SFT stage, which controls the relative contribution of the reflection and correction objective to the overall training signal; the coefficient \(\beta_{cor}\) in the RL stage, which scales the overall strength of the reflection and correction reward against the task reward \(R_{task}\); the factor \(\beta_{last}\),  which controls the trade-off between the two components of \(R_{task}\), thereby emphasizing the quality of the final corrected output; and the weights \(\beta_{loc}\) and \(\beta_{sem}\) inside \(R_{cor}\), which govern the impact of position-level and bucket-level improvement signals when shaping the correction reward. These experiments are conducted on the industrial dataset, using five perturbed values in the neighborhood of the chosen value. Figure \ref{fig:sensitive_v1} summarizes the results of these hyperparameter tuning experiments, from which we make the following observations.
\begin{itemize}[noitemsep, topsep=0pt, leftmargin=*]
\item \textbf{Correction Template coefficient \(\lambda_{rc}\)}: GRC maintains high and stable performance as \(\lambda_{rc}\) varies from 0.8 to 1.6, with only a marginal improvement around 1.0–1.2, indicating that once the subsequent RL stage is applied, the model becomes largely insensitive to the SFT loss weight.
\item \textbf{Correction Reward coefficient \(\beta_{cor}\)}: GRC consistently improves across all recall metrics as \(\beta_{cor}\) increases from 1.4 to 2.2, suggesting that strengthening the reflection–correction rewards in the RL stage effectively exploits the enlarged decision space and yields substantial performance gains.
\item \textbf{Token Reward coefficient \(\beta_{last}\)}: GRC benefits steadily from increasing \(\beta_{last}\) in the range 1.0–2.0, showing that placing more emphasis on the reward of post-reflection tokens (relative to pre-reflection tokens) leads to better retrieval quality.
\item \textbf{Position and Semantic Correction Coefficients $\beta_{loc}$, $\beta_{sem}$}: GRC remains robust when $\beta_{loc}$ ranges from 0.8 to 1.4 and $\beta_{sem}$ ranges from 0.6 to 1.2, with only slight performance degradation observed at $\beta_{loc}=0.6$ and $\beta_{sem}=0.4$. This suggests that the model is generally insensitive to both token- and semantic-level correction rewards unless they are excessively down-weighted.
\item \textbf{Entropy Scheduling Coefficient \(\alpha_{e}\)}: GRC exhibits moderate sensitivity to \(\alpha_{e}\), with Recall@100/200 peaking around \(\alpha_{e}=0.2\). When \(\alpha_{e}\) is too large, the entropy term overly dominates the beam score and suppresses high-quality token combinations favored by standard beam search; when it is too small, low-entropy (high-confidence) trajectories occupy more beam slots, leaving insufficient capacity for uncertain paths that contribute to tail recall. These results indicate that a properly tuned entropy weight is essential for fully exploiting entropy-guided beam search.
\end{itemize}

\section{Online Experiments}
We further evaluated GRC in an online A/B test on our e-commerce advertising platform from Jan 2 to Jan 12, 2026. We used our in-house TIGER implementation as the baseline and deployed GRC as the treatment. The control and treatment buckets each contained 15\% of randomly sampled users. Compared with TIGER, GRC improved advertising revenue by \textbf{1.79\%}, CTR by \textbf{2.11\%}, and GMV by \textbf{2.04\%}, all statistically significant ($p < 0.05$, two-sided test). With FlashAttention~\cite{flashattention} enabled for both variants to accelerate inference, P99 latency increased modestly from 27\,ms to 31\,ms due to the additional reflection and correction passes. Overall, GRC provides a favorable accuracy--latency trade-off in production.

\vspace{-0.8em}
\section{Conclusion}
In this paper, we propose GRC, a structured reflection-correction framework for generative recommendation that mitigates error accumulation of one-pass decoding. GRC explicitly extends standard decoding into a Generation–Reflection–Correction process in the semantic-token space, enabling the model to detect deviations in the first draft and revise trajectories conditioned on structured, multi-granular reflection signals. To better optimize reflection–correction decisions in the enlarged trajectory space, we further introduce a GRPO-based reinforcement learning stage with well-designed rewards that jointly capture token accuracy, error localization, semantic consistency, and post-correction improvement. During inference, we design EGRS, an entropy-guided scheduling strategy integrated with beam search, to allocate limited correction budget to high-uncertainty trajectories under latency constraints. Extensive offline experiments and online A/B tests in a large-scale commercial system demonstrate that GRC consistently improves retrieval quality with modest latency overhead, offering a practical and scalable approach to self-corrective generative recommendation.
We believe this framework helps bridge the gap between one-pass GR and robust, reasoning-aware recommenders, and can serve as a foundation for future work on reflection-enhanced GR.

\bibliographystyle{ACM-Reference-Format}
\bibliography{reference}
\end{document}